\documentclass[runningheads]{llncs}
\usepackage{graphicx}
\usepackage{comment}
\usepackage{amsmath,amssymb} 
\usepackage{color}
\usepackage{floatrow}
\usepackage{multirow}
\newfloatcommand{capbtabbox}{table}[][\FBwidth]
\usepackage[utf8]{inputenc}
\usepackage{blindtext}

\begin{document}
\pagestyle{headings}
\mainmatter

\title{DR-KFS: A Differentiable Visual Similarity Metric for 3D Shape Reconstruction} 

\titlerunning{DR-KFS: A Differentiable Visual Similarity Metric for 3D Shape Recon.}
\author{Jiongchao Jin\inst{1,2} \and
Akshay Gadi Patil\inst{1} \and
Zhang Xiong\inst{2} \and
Hao Zhang\inst{1}}
\authorrunning{Jin et al.}
%
\institute{Simon Fraser University \and
Beihang University}
\maketitle

\vspace{-10pt}

\begin{figure}[] \centering
	\includegraphics[width=0.99\linewidth]{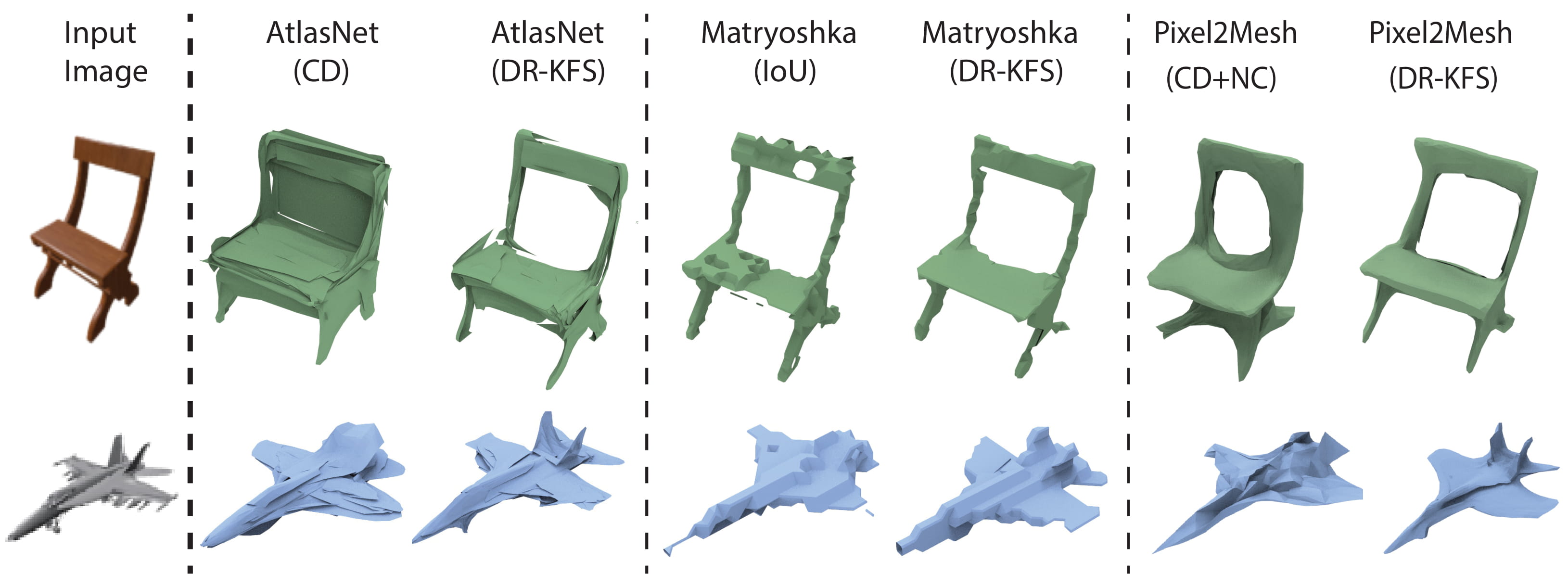}
	\caption{Single-view 3D reconstruction by AtlasNet~\cite{atlasnet}, Matryoshka Network~\cite{Matryoshka}, and Pixel2Mesh \cite{pixel2mesh}, trained with different losses. Left of each pair: trained by shape distortion in CD, IoU, and a combination of CD and normal consistency (NC). Right of each pair: using DR-KFS, our {\em differentiable visual similarity\/} metric, as train loss leads to results of higher visual fidelity in terms of shape structures and surface quality.}
	\label{fig:teaser}
\end{figure}

\vspace{-20pt}

\begin{abstract}
We introduce a {\em differential visual similarity\/} metric to train deep neural networks for 3D reconstruction, aimed at improving reconstruction quality. The metric compares two 3D shapes by measuring distances between multi-view images {\em differentiably\/} rendered from the shapes. Importantly, the image-space distance is also differentiable and measures visual {\em similarity\/}, rather than pixel-wise distortion. Specifically, the similarity is defined by mean-squared errors over HardNet features computed from probabilistic keypoint maps of the compared images. 
Our differential visual shape similarity metric can be easily plugged into various 3D reconstruction networks, 
replacing their distortion-based losses,
such as Chamfer or Earth Mover distances, so as 
to optimize the network weights to produce reconstructions with better structural fidelity and visual quality. We demonstrate this both objectively, using well-known shape metrics for retrieval and classification tasks that are independent from our new metric, and subjectively 
through a perceptual study.   
\end{abstract}


\section{Introduction}
\label{sec:intro}

Reconstructing 3D structures from 2D images is one of the most fundamental problems in computer vision. The problem is clearly ill-posed, hence most reconstruction algorithms are expected to incur errors against the ground truth. Commonly used error metrics are defined by {\em shape distortions}, such as Chamfer Distance (CD), Earth Mover Distance (EMD), Mean Square Error (MSE), and Intersection over Union (IoU). Most deep neural networks developed for 3D reconstruction~\cite{atlasnet,PointOutNet,OGN,hspHane17,pixel2mesh,PTN,softras}
are trained by loss functions defined by one of these shape distortion metrics. 
To serve as a network loss to enable back propagation, the adopted metric needs to be {\em differentiable\/}. All of CD, EMD, and MSE are differentiable. While the original IoU is not, there have been recent attempts to make it differentiable~\cite{yu2016,softras}.

%
%

\begin{figure}[!t] \centering
	\includegraphics[width=0.75\linewidth]{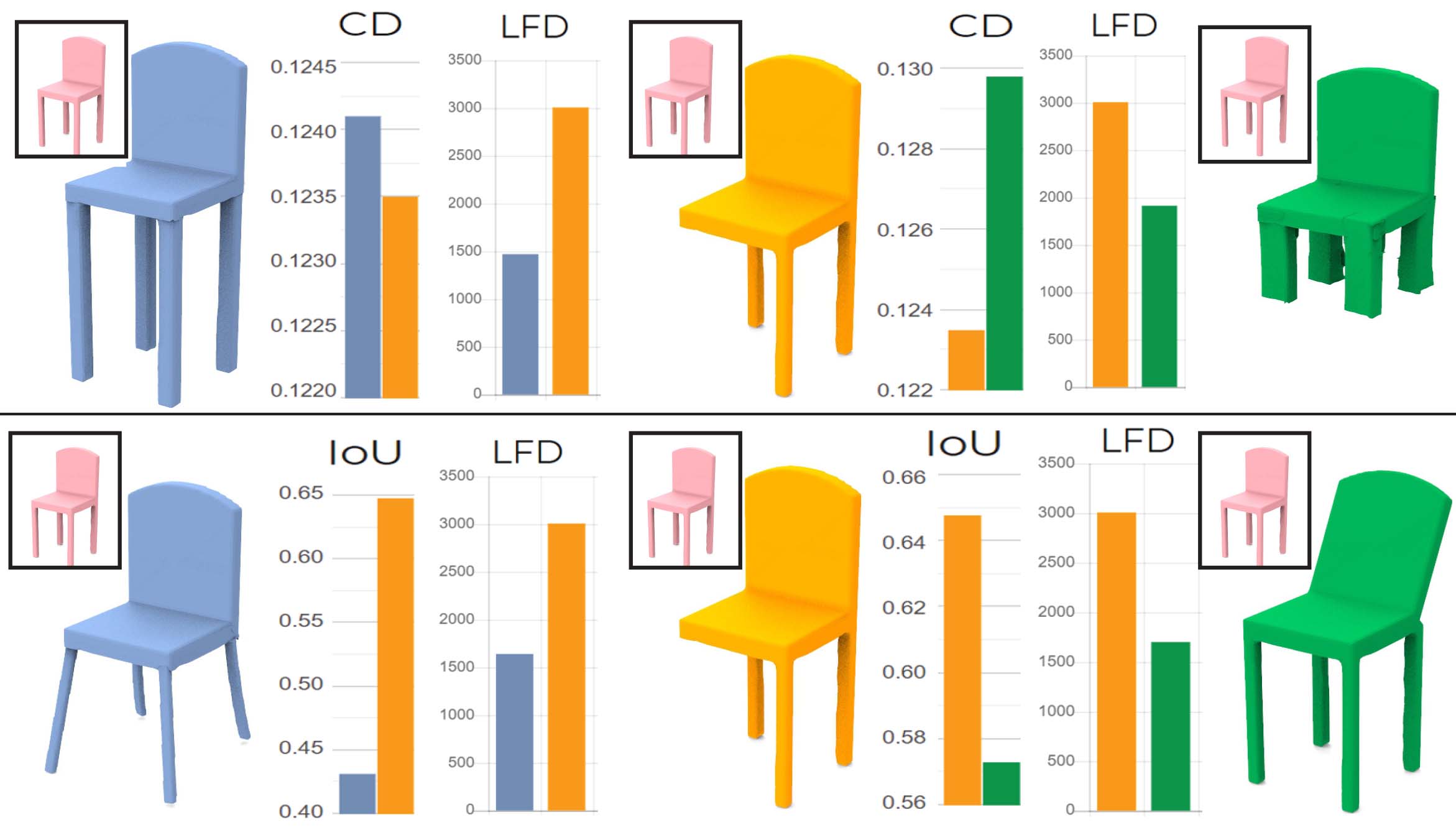}
	\caption{A {\em trade-off\/} between shape distortion measures, e.g., CD and IoU, and a visual similarity metric such as LFD.
		Top: changing the shapes of the legs of a chair, e.g., by thickening and elongating (blue) or shortening (green), leads to larger shape distortions (reflected by larger CDs) compared to removing one leg (orange). The leg removal appears to make more of a visual impact due to a {\em structure\/} alteration and it is captured by a larger LFD (orange bars). Note that the bars show CDs/LFDs between the changed chair and the original, shown in pink in the boxes. Bottom: bending the legs (blue) or the back (green) leads to larger shape distortion, as reflected by {\em smaller\/} IoU, compared to leg removal. Again, the latter is captured by a larger LFD.
	}
    \vspace{-5pt}
	\label{fig:cd_lfd}
\end{figure}

What is common about these well-adopted distortion metrics is that they are all designed to measure geometric distances between {\em aligned\/} 3D shapes, in the {\em object space\/} --- they do not account for how the shapes are viewed by human observers, i.e., they are not ``visual''.
Many recent works~\cite{IMGAN,OccNet,Tatarchenko2019,pontes2018image2mesh,PointOutNet,Matryoshka} have shown relative insensitivity of these object-space distortion metrics to {\em structural errors\/} such as missing parts and topological noise, and visual artifacts such as self intersections and poor surface quality. For example, thickening and elongating/shortening all four legs of a chair may result in a larger CD than removing one of the legs entirely, yet the latter alteration, a structural change, is more visually apparent; see Figure~\ref{fig:cd_lfd}. As a result, while some reconstructed 3D shapes do exhibit better visual quality, ratings based on distortion measures such as CD, EMD, MSE, or IoU may not reflect that superiority. 
This is not entirely surprising — a recent work by Blau and Michaeli~\cite{blau2018perception} even suggests a trade-off between perceptual and distortion measures, albeit for image restoration.

Our key observation is that in contrast to shape distortion, measures of {\em visual similarity\/} are less sensitive to misalignment and slight shape distortion, but  more sensitive to structural errors and visual quality.
In computer graphics, 
the best known visual similarity metric 
is the light field descriptor (LFD)~\cite{LFD}. LFDs are computed for silhouette images of 3D shapes rendered from 
multiple viewpoints sampled around the shapes. Both a contour-based (Fourier descriptor) 
and a region-based (Zernike moment) image-space descriptor are employed, where rotational alignment is 
resolved via a discrete exhaustive search. However, due to the discrete rasterization during rendering and the 
use of truncated Fourier descriptors, LFD is non-differentiable.
%
Some works in computer vision have also utilized multi-view, {\em projective\/} losses to train reconstruction networks, e.g., Perspective Transformer Net (PTN)~\cite{PTN}, Neural Mesh Render~\cite{NMR}, and SoftRas~\cite{softras}. However, the projective- or image-space distances employed by these methods are still based on pixel-wise distortion, rather than similarity.

In this paper, we advocate the use of {\em differentiable visual shape similarity\/} metrics to train deep neural networks (DNNs) for 3D reconstruction. We introduce such a metric which compares two 3D shapes
by measuring visual, image-space 
similarity
between multi-view images rendered from the shapes, similar
to LFD. However, one key difference is that the rendering process is differentiable, by employing a simplified {\em soft rasterization\/}~\cite{softras}. In addition, we develop a differentiable similarity distance in the image space based on MSE defined over {\em probabilistic keypoint maps\/} of the compared images, rather than on RGB values. Furthermore, the MSE is defined over HardNet~\cite{HardNet} features, rather than on the original keypoint maps. 
This choice is motivated in part by the finding in~\cite{blau2018perception} that the perception-distortion trade-off appears
less severe for distance between VGG features.
Putting all these together, we arrive at a differentiable visual similarity metric, which we call DR-KFS to capture the use of Differentiable Rendering, Keypoint maps, and Feature-space image Similarity distances. DR-KFS better matches visual evaluation by humans, compared to object-space metrics.

To the best of our knowledge, existing 3D reconstruction networks all employ 
distortion-based
reconstruction losses, 
either in the object or image spaces
see Tables~\ref{tab:big_list} and~\ref{tab:Visual distortion loss} for a summary. Our differential visual 
similarity
metric can be easily plugged into these networks, 
replacing the distortion losses so as to optimize the network weights to produce reconstruction results 
with better structural fidelity and visual quality. We demonstrate this both objectively, using well-known visual shape metrics for retrieval and classification tasks that are {\em independent\/} from DR-KFS, and subjectively through a user study. 

Specifically, the 3D reconstruction networks tested for adaptation to DR-KFS training include OGN, AtlasNet~\cite{atlasnet}, and Matryoshka Networks~\cite{Matryoshka}, which were picked as representative networks by Tatarchenko et al.~\cite{Tatarchenko2019} in their recent systematic study of single-view 3D reconstruction. We also consider Pixel2Mesh~\cite{pixel2mesh}, 
NMR~\cite{NMR}, SoftRas\cite{softras}
and 3D-R2N2~\cite{3DR2N2}. The visual shape metrics we employ for evaluation include Shape Google~\cite{ShapeGoogle}, Multiview CNN or MVCNN~\cite{mvcnn}, normal consistency or NC~\cite{OccNet}, F-score~\cite{Tatarchenko2019,Matryoshka}, as well as LFD. It is worth noting that the last three metrics have all been adopted by recent works on 3D reconstruction~\cite{OccNet,Tatarchenko2019,Matryoshka,IMGAN} to complement the distortion metrics.


\section{Related work}
\label{sec:related}


\begin{table*}[t!]%
	\begin{center}
		\begin{tabular}{p{0.25\textwidth}|p{0.22\textwidth}|p{0.23\textwidth}|p{0.25\textwidth}}
			\hline
			{\bf Method\/} &{\bf Representation\/}& {\bf Training loss\/} & {\bf Evaluation metrics\/} \\
			\hline\hline
			AtlasNet~\cite{atlasnet} &Points + mesh\/& CD &CD \\ \hline
			PointOutNet~\cite{PointOutNet} & Points & CD, EMD &CD, EMD \\ \hline
			OGN~\cite{OGN} &Octree& cross-entropy/MSE &CD, EMD \\ \hline
			HSP~\cite{hspHane17} &Voxel& cross-entropy/MSE&CD, IoU \\ \hline
			Pixel2Mesh~\cite{pixel2mesh} &Points + mesh& CD, surface normal & CD, EMD\\ \hline
			Matryoshka Net~\cite{Matryoshka} & Voxel shape layer & IoU, cos-similarity& IoU \\ \hline
			IM-Net~\cite{IMGAN} & Implicit field & Per-point status &CD, MSE, IoU, LFD \\ \hline
			OccNet~\cite{OccNet} &Implicit field& Per-point status &CD, IoU, NC \\ \hline
			DeepSDF~\cite{deepSDF} & Implicit field & Per-point status&CD, EMD \\ \hline
			DISN~\cite{DISN} & Implicit field & Per-point status &CD, F-score, IoU\\ \hline
		\end{tabular}
		\vspace{-5pt}
		\caption{\textit{Object-space training losses and evaluation metrics for state-of-the-art 3D reconstruction networks.} CD and EMD are widely adopted in point-based networks, while MSE, cross-entropy, and differentially implemented IoU are more popular in voxel-based reconstruction. Implicit decoders all use per-point reconstruction loss.}
		\label{tab:big_list}
	\end{center}
\end{table*}

\begin{table*}[t!]%
	\begin{center}
		\begin{tabular}{p{0.20\textwidth}|p{0.35\textwidth}|p{0.21\textwidth}|p{0.20\textwidth}}
			\hline
			{\bf Method\/} &{\bf 2D observation\/}& {\bf Training loss\/} &{\bf{Evaluation}}\\
			\hline\hline
			PTN~\cite{PTN} &Silhouettes \/& Pixel-wised MSE &IoU \\ \hline
			NMR~\cite{NMR} & Silhouettes & Pixel-wised MSE &IoU\\ \hline
			Soltani et al.~\cite{synthesizing} & Depth Maps and Silhouettes &L1 norm &IoU \\ \hline
			SoftRas~\cite{softras} & RGBA images & IoU & IoU \\ \hline
		\end{tabular}
		\vspace{-10pt}
		\caption{\textit{Multi-view training losses and evaluation metrics for state-of-the-art 3D reconstruction networks.} All the image-space errors are based on pixel-wise distortion such as MSE, IoU, or L1 Norm. All the evaluations are based on IoU.}
		\label{tab:Visual distortion loss}
	\end{center}
\end{table*}

\paragraph{DNNs for 3D reconstruction.}
%
One of the earlier DNNs for 3D reconstruction is 3D-R2N2~\cite{3DR2N2}, which proposes the use of Recurrent Neural Networks (RNNs) to reconstruct a voxelized shape from a single-view image. Another notable work, PointOutNet \cite{PointOutNet}, produces multiple reconstruction candidates for a single-view input image and demonstrates an improvement in the reconstruction quality over 3D-R2N2. OGN \cite{OGN} generates volumetric 3D outputs in a compute-and-memory efficient manner by using an octree representation. AtlasNet \cite{atlasnet} uses a collection of parametric surface elements to represent a 3D shape and to naturally infer a surface representation of the shape. Pixel2Mesh \cite{pixel2mesh} produces 3D triangle meshes from a single image using Graph Convolution Networks (GCN) \cite{GCNs}. Matryoshka Network \cite{Matryoshka} introduces a novel and efficient 2D encoding scheme for 3D geometry, posing 3D reconstruction as a 2D prediction problem, while also speeding up the process.

Recent works on generative shape modeling using implicit functions have achieved state-of-the-art visual quality. IM-NET \cite{IMGAN} and OccNet \cite{OccNet} represent 3D surfaces implicitly in the form of continuous binary decision functions. DeepSDF \cite{deepSDF} learns a continuous Signed Distance Function (SDF) representation for a 3D shape. DISN \cite{DISN} combines implicit SDF descriptors with the local input image features to generate 3D shape surfaces. All these implicit function-based networks can generate 3D models with a smooth surface and good visual quality. Despite the good visual quality of the generated surfaces, they often obtain lower scores on 
\emph{shape distortion}
metrics, such as CD.

\vspace{-7pt}

\paragraph{Object-space metrics for 3D reconstruction.}
%

Table~\ref{tab:score_table} lists state-of-the-art 3D reconstruction DNNs trained using object-space losses,
along with the shape metrics they employed. In principle, a network that is trained on a particular loss, e.g., CD, should be expected to perform better on that metric during testing, than methods that were not trained with the metric. 
Most
methods that we are aware of train their networks using object-space distortion metrics. CD and EMD, and to a lesser degree, IoU, are the dominant measures applied for evaluating reconstruction quality. It is interesting to observe that some of the most recent methods, including IM-Net~\cite{IMGAN}, OccNet~\cite{OccNet}, and Matryoshka Network~\cite{Matryoshka}, have opted to evaluate their methods using alternative measures such as LFD, Normal Consistency (NC), and F-score. Not surprisingly, each of these works pointed out the shortcomings of object-space distortion metrics in capturing visual similarity, which motivated their use of other alternatives. In terms of visual quality of the 3D reconstruction, the current state-of-the-art results are obtained by methods based on learning implicit fields~\cite{IMGAN,OccNet,deepSDF,DISN}.

\vspace{-7pt}

\paragraph{Visual shape similarity measures.}
For shape retrieval and classification, many visual shape similarity measures have been proposed. Here we have chosen a subset of them as representative measures. 
LFD~\cite{LFD} performs visual shape similarity by extracting local features from one-hundred orthogonal projections of each 3D model. 
ShapeGoogle~\cite{ShapeGoogle} constructs compact and informative shape descriptor using a \emph{bag-of-visual- features}. 
MVCNN~\cite{mvcnn} combines information from multiple views of a 3D shape into a single and compact shape descriptor with the help of CNNs. 
F-score~\cite{Tatarchenko2019} performs visual shape similarity explicitly by calculating the distance between object surfaces using a harmonic mean of precision and recall scores. 
And finally, normal consistency (NC)~\cite{OccNet,pixel2mesh} measures how well any 3D reconstruction can capture higher order information by calculating a mean absolute dot product of face-normals of the given 3D models, represented as meshes. 
However, LFD, ShapeGoogle, F-score are not differentiable, and NC measure  can only capture higher-order information.

\vspace{-7pt}

\paragraph{Image-space similarity.}
%
Most common quantitative measures of image similarities include pixel-wise MSE and IoU. However, such measures are not designed to compare visual similarity between objects captured in the images, e.g., they are sensitive to misalignment.
Hand-crafted image descriptors such as SIFT \cite{SIFT}, SURF \cite{SURF}, and ORB \cite{ORB} can be perform image retrieval to visually correspond closest image matches, but they are discrete and non-differentiable. On the other hand, the use of powerful image descriptors obtained from CNNs has been shown to make the image retrieval pipeline differentiable \cite{LIFT,AffNet,Ppfnet,ConDes,mvcnn}, allowing an end-to-end neural network approach for measuring image similarity.

Generic image similarity computation has to account for rotation, translation, and occlusion of identical entities in the two images, as well as the overall illumination variation. However, the 3D models input to our framework are aligned and consequently, the rendered view-images are aligned as well, and as such do not pose the above challenges.
Our inspiration to develop a differentiable and visual image-similarity metric for rendered images of 3D shapes stems from the development of differentiable and visual image-retrieval CNNs.




\vspace{-7pt}

\paragraph{Multi-view/projective approaches to 3D reconstruction.}
%
Some works on single-view 3D reconstruction take a multi-view approach to train their DNNs. They rely on either a differentiable renderer or multi-view projections of 3D shapes and perform supervised learning on the obtained images rather than in the 3D object space, e.g., PTN~\cite{PTN}. Table~\ref{tab:Visual distortion loss} lists notable recent works along these lines and the losses and metrics they employed for training and evaluation. Specifically, 
NMR \cite{NMR}, which outperformed PTN, reconstructs a 3D mesh from a single image with silhouette-image supervision. Other works such as MarrNet~\cite{marrnet} and Soltani et al.~\cite{synthesizing} use 2.5D sketch images and rendered silhouette images, respectively, for learning single-view 3D reconstruction. Most recently, SoftRas~\cite{softras} reconstructs a 3D shape using shaded images, i.e., view-renderings of the input RGB image with a light source. Importantly, SoftRas converts the discrete raster operation to a probabilistic soft raster one, directly solving the indifferentiable raster operation of the traditional renderer present in \cite{PTN,NMR}. However, in all these works, the image-space errors are still measured by distortion losses such as pixel-wise MSE or IoU, which do not capture shape similarity. 
%
%
In contrast, our metric, DR-KFS, is designed to capture visual similarity of the captured shapes.
%
The renderer used in DR-KFS is a simplified version of SoftRas~\cite{softras}, which is more efficient than CNN-based methods for object rendering.

\section{Differential visual shape similarity metric}
\label{sec:method}

\begin{figure*}[t] 
	\centering
	\includegraphics[width=\linewidth]{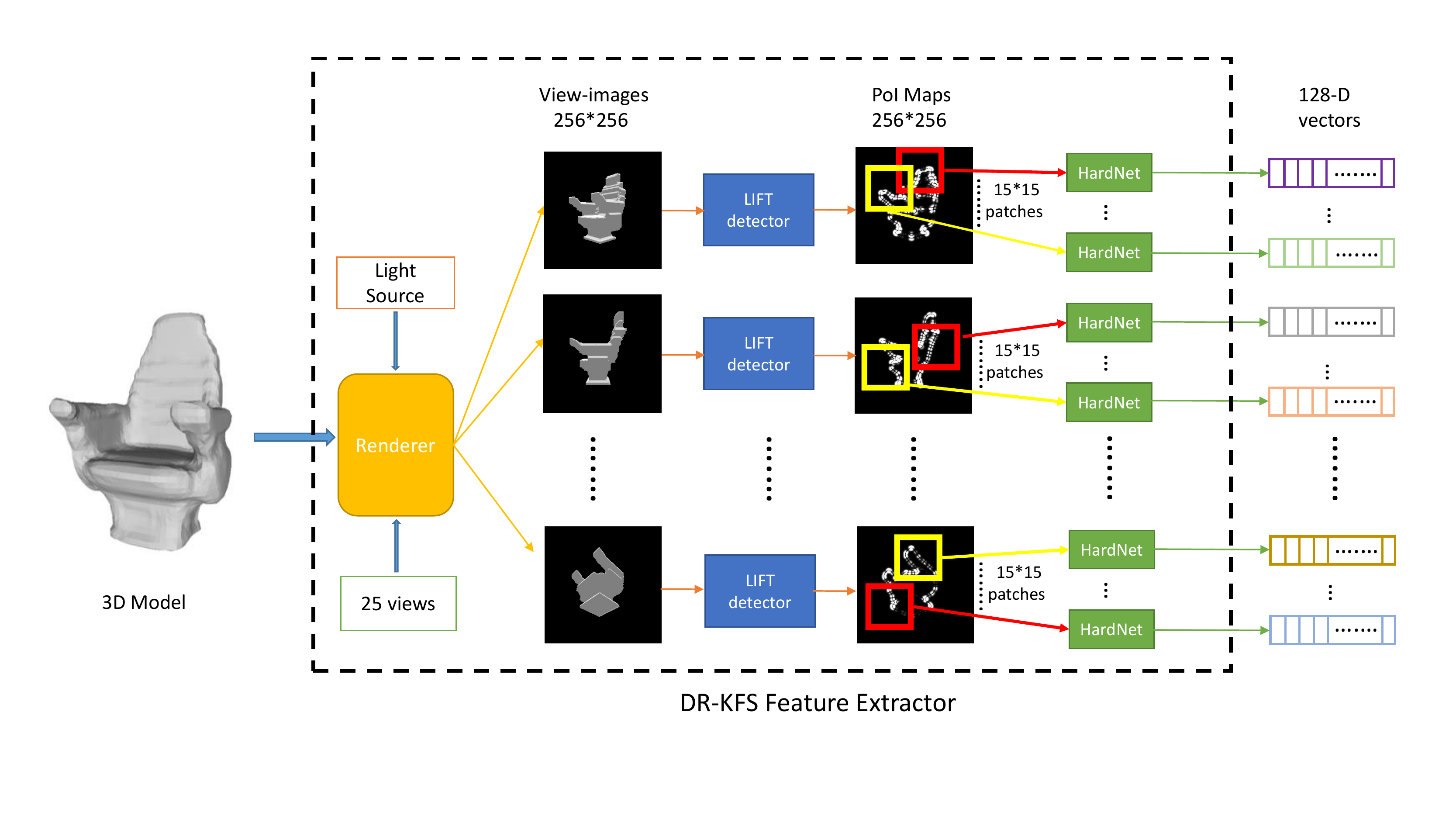}
	\vspace{-3em}
	\caption{Overall framework for obtaining DR-KFS descriptors using continuous patch features. The first step is the differentiable renderer, a \emph{Soft Rasterizer}, which renders multi-view images. This renderer, coupled with the process of continuous patch feature extraction over the point-of-interest (PoI) maps, makes our framework differentiable.}
	\label{fig:pipeline_a}
\end{figure*}

Visual shape similarity metrics, such as LFD \cite{LFD}, rely on projections or renderings of multi-view images of 3D models from a set of view angles. We follow this theme of render-and-match-images to compare the visual similarity between two 3D shapes while make the framework {\em differentiable\/} so that it can be easily plugged into any 3D reconstruction DNN; see Figure \ref{fig:pipeline_b} for  an overall pipeline.

Given two 3D models, one reconstructed using an existing approach such as AtlasNet \cite{atlasnet}, Matryoshka Network \cite{Matryoshka}, Pixel2Mesh \cite{pixel2mesh}, or OGN \cite{OGN}, and the other being the corresponding ground truth (GT) model, we first align them and render the models from twenty-five viewing angles, using the \emph{Soft Rasterizer} \cite{softras} renderer, which is differentiable. For each view-image, a Point of Interest (PoI) map is obtained using a keypoint detection network. Visual similarity of the 3D models is determined using DR-KFS feature matching (see Figure \ref{fig:pipeline_b}) obtained by extracting local features from PoI maps. 


%
\subsection{Differentiable renderer}
\label{subsec:diff_rend}

To measure 3D shape similarity using LFD, projected images from many angles are required. Due to hardware constraints in training our DNNs, we only use 25 viewpoints, sampled on a semi-sphere as an approximation to the ideal case. To achieve differentiability in the rendering process, we replace rasterization and z-buffering operations used in conventional rendering pipelines with soft rasterization and probabilistic map aggregation \cite{softras}. No color information is used in DR-KFS renderer as many reconstruction networks are incompatible with texture projections on the reconstructed model surfaces. A fixed position light source is aptly placed to provide information about shape surface quality when rendering the view-images. We make use of the rendered view-images to determine the similarity of given 3D models, as explained below.

\begin{figure*}[t]
	\centering
	\includegraphics[width=\linewidth]{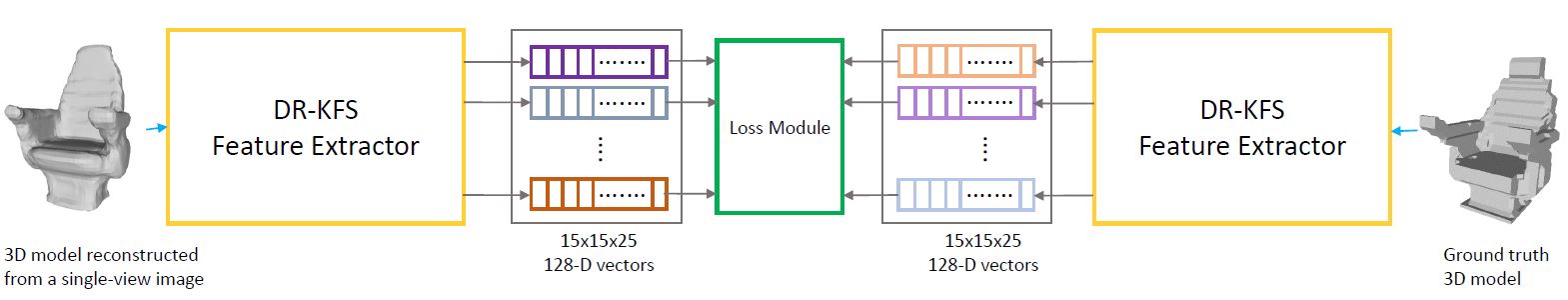}
	\vspace{-5 pt}
	\caption{An end-to-end pipeline of our  visual, differentiable shape-similarity metric framework for single view 3D reconstruction. Feature descriptors from Figure \ref{fig:pipeline_a} are used to calculate the overall loss, which is backpropagated to the 3D reconstruction network (which reconstructs a 3D model, shown on the left) via the associated DR-KFS module, improving the reconstruction quality.}
	\label{fig:pipeline_b}
	\vspace{-1 em}
\end{figure*}

%
Our visual similarity metric is based on matching rendered view-images by extracting discriminative local features over Point of Interest (PoI) maps, using Convolutional Neural Networks (CNNs).

\paragraph{PoI map.}
A PoI map is a probabilistic map that gives a score for every pixel being a keypoint. It is generated by a CNN-based keypoint detection network corresponding to every view-image. To this end, we borrow the LIFT detector introduced in \cite{LIFT}, which uses piece-wise linear activation functions \cite{TILDE} in convolution layers to get a PoI map for an input image. It is formulated as:
\begin{equation}
\label{equa:PoI}
PoI_{map} = f_{Net}(I) = \sum_{n}^{N}\delta_{n} \max_{m}^{M} (W_{mn} \circledast I + b_{mn})
\end{equation}
where $f_{Net}(I)$ is a non-linear function of the rendered view-image \emph{I}, using a neural network $Net$, which is nothing but a CNN-based keypoint detector. $N,M$ are hyperparameters controlling the complexity of the piece-wise linear activation function. $\delta$ is +1 if $n$ is odd, and -1 otherwise. The parameters of the network $Net$ to be learned are the convolution filter weights $W_{mn}$ and biases $b_{mn}$, and \emph{$\circledast$} denotes the convolution operation. A detailed description of LIFT based keypoint detection can be found in \cite{LIFT}.
%

\paragraph{Continuous patch features.}
After generating PoI maps for each of the rendered images, local patch features are extracted from them using a sliding window of size $32\times32$ with a stride of $16$, resulting in 225 ($15\times15$) patches per image $I$. These patches span the entire image and there exists an overlap between adjacent patches which provides continuity over the feature space. 

Note that in existing image retrieval works, such as in \cite{LIFT,TILDE}, detected keypoints are projected back onto the original image and image features are extracted locally around these keypoints. Image retrieval is performed by matching these keypoint-based local features. This is a \emph{discrete} process over the image space. As a result, this approach of matching images based on discrete keypoints, if employed, introduces non-differentiability into the latter parts. Unlike such methods, we avail continuous local patch features from the PoI maps by employing a sliding window technique spanning the entire map. Our framework is inspired by the recent work of \cite{blau2018perception} which shows that local patch features help alleviate the problem of ``human perception $vs$ image distortion" trade-off.

A sliding window on a PoI map, followed by the HardNet\cite{HardNet} feature extractor outputs a 128-D feature descriptor per patch (see Figure \ref{fig:pipeline_a}).
The HardNet \cite{HardNet} module we use is pretrained on UBC PhotoTour\cite{PhotoTour}, a standard dataset used to extract local patch features. We coin these features as the DR-KFS feature descriptors. The continuous patch features, coupled with the early-stage render (\emph{Soft Rasterizer}) allow our pipeline to be entirely differentiable, demonstrated further by our end-to-end learning framework in Figure \ref{fig:pipeline_b}.
%

\paragraph{Feature matching.} 
After extracting the DR-KFS feature descriptors for the two 3D models, we perform feature matching. For every DR-KFS feature descriptor of the reconstructed 3D model, we find the best DR-KFS feature descriptor of the GT 3D model, in terms of minimum mean squared error (MSE) loss. We do not correspond patches merely based on their position in the PoI maps. This can be thought of as a feature matching function. Symbolically, this function finds the best match for every PoI patch of the reconstructed model, from the PoI patches of the ground truth model.

After feature matching, the total loss is formulated as the linear sum of the pairwise (best match) MSE loss, 
which is backpropagated through the DR-KFS feature descriptor module associated with the reconstructed 3D model (Fig \ref{fig:pipeline_b}). 




\section{Results and Evaluation}
\label{sec:exp}

\begin{table*}[t!]%
	\begin{center}
		\begin{tabular}{p{0.155\textwidth}|p{0.138\textwidth}|p{0.07\textwidth}|p{0.08\textwidth}|p{0.065\textwidth}|p{0.08\textwidth}|p{0.08\textwidth}|p{0.07\textwidth}|p{0.08\textwidth}|p{0.08\textwidth}}	
			\hline
			\hline
			{Networks}&{Training}&
			\multicolumn{7}{c}{\ \ \ \ \ \ Evaluation Metrics}\\
			\cline{3-10}
			
			{}&{Strategies} &{CD\/}& {IoU\/} & {SG\/} &{FS}&{NC}&{LFD}&{{\tiny MVCNN}}&{{\tiny DR-KFS}}\\
			\hline
			\bf {Atlas25\cite{atlasnet}}
			&Original &\bf{6.53} &53.38 &3.17&63.17&\bf{0.792} &4338&0.532&0.288\\ 
			&DR-KFS~ &7.11   &\bf{54.41}&\bf{2.26}&\bf{65.4}8&0.790&\bf{3796}&\bf{0.522}&\bf{0.213}\\
			\hline
			{\bf MN\cite{Matryoshka}}
			&Original~&\bf{2.86} &65.33&4.32&\bf{74.22}&0.825&3866&\bf{0.637}&0.300\\ 
			&DR-KFS ~ & 2.91&\bf{66.08}&\bf{2.08}&65.30    &\bf{0.829}&\bf{3675}&0.660&\bf{0.265}\\
			\hline
			{\bf P2M\cite{pixel2mesh}}
			&Original~&\bf{6.61}&54.08&6.08&58.15&\bf{0.761}&4508& \bf{0.648}&0.463\\ 
			&DR-KFS ~      &6.98&\bf{54.32}&\bf{3.84}&\bf{60.23}&0.754&\bf{4212}&0.662&\bf{0.332}\\
			\hline
			{\bf OGN\cite{OGN}}
			&Original~ & 6.13&\bf{57.01}&9.94&\bf{56.36}&0.695&4436&0.752&0.522\\ 
			&DR-KFS ~&\bf{6.08}&56.83&\bf{8.73}&55.09&\bf{0.702}&\bf{4237}&\bf{0.661}&\bf{0.461}\\
			\hline
			{\bf 3D-}
			&Original~ & \bf{7.42}&\bf{53.98}&9.75&48.09&0.621&4692&0.841&0.539\\ 
			{\bf{R2N2\cite{3DR2N2}}}&DR-KFS ~& 7.56 &52.74&\bf{9.13}&{\bf{51.58}}&{\bf{0.681}}&\bf{4416}&\bf{0.713}&\bf{0.441}\\
			\hline
			{\bf Shape}
			&NMR\cite{NMR}~ & 11.95&48.90&8.13& 51.72&0.522 &7716&0.842&0.560\\ 
			{\bf Generator}&SoftRas\cite{softras} ~& 9.75&54.53&7.11&57.08&0.685&5385&\bf{0.631}&0.471\\
			{\bf{\cite{NMR,softras}}}&DR-KFS ~& \bf{8.06}&\bf{55.24}&\bf{5.48}&\bf{62.02}&\bf{0.634}&\bf{4125}&0.741&\bf{0.395}\\
			\hline
			{\bf IM-Net\cite{IMGAN}}
			&Original~ &7.02 & 66.01&1.38&69.43&0.740&3806 &0.511&0.203\\ 
			
			\hline			
		\end{tabular}	
	\end{center}
	\vspace{-1 em}
	\caption{Quantitative comparison for single-view 3D reconstruction networks when trained using their original distortion-based losses vs.~DR-KFS, our differentiable visual similarity metric. Tested DNNs include AtlasNet25~\cite{atlasnet} (Atlas25), Matryoshka Net~\cite{Matryoshka} (MN), Pixel2Mesh~\cite{pixel2mesh} (P2M), Octree Generating Network~\cite{OGN} (OGN), 3D-R2N2~\cite{3DR2N2}, and shape generation using NMR~\cite{NMR} and SoftRas~\cite{softras}. Evaluation is done on both object-space (CD and IoU) and visual similarity metrics: SG = Shape Google, FS = F score, NC = Normal Consistency, LFD = Light Field Descriptor, {\tiny MVCNN\/} = multi-view CNN. Numbers reported represent average performance over four object categories.}
	\label{tab:score_table}
\end{table*}

\begin{figure*}
	\centering
	{\includegraphics[width=\linewidth, height=1.18\linewidth]{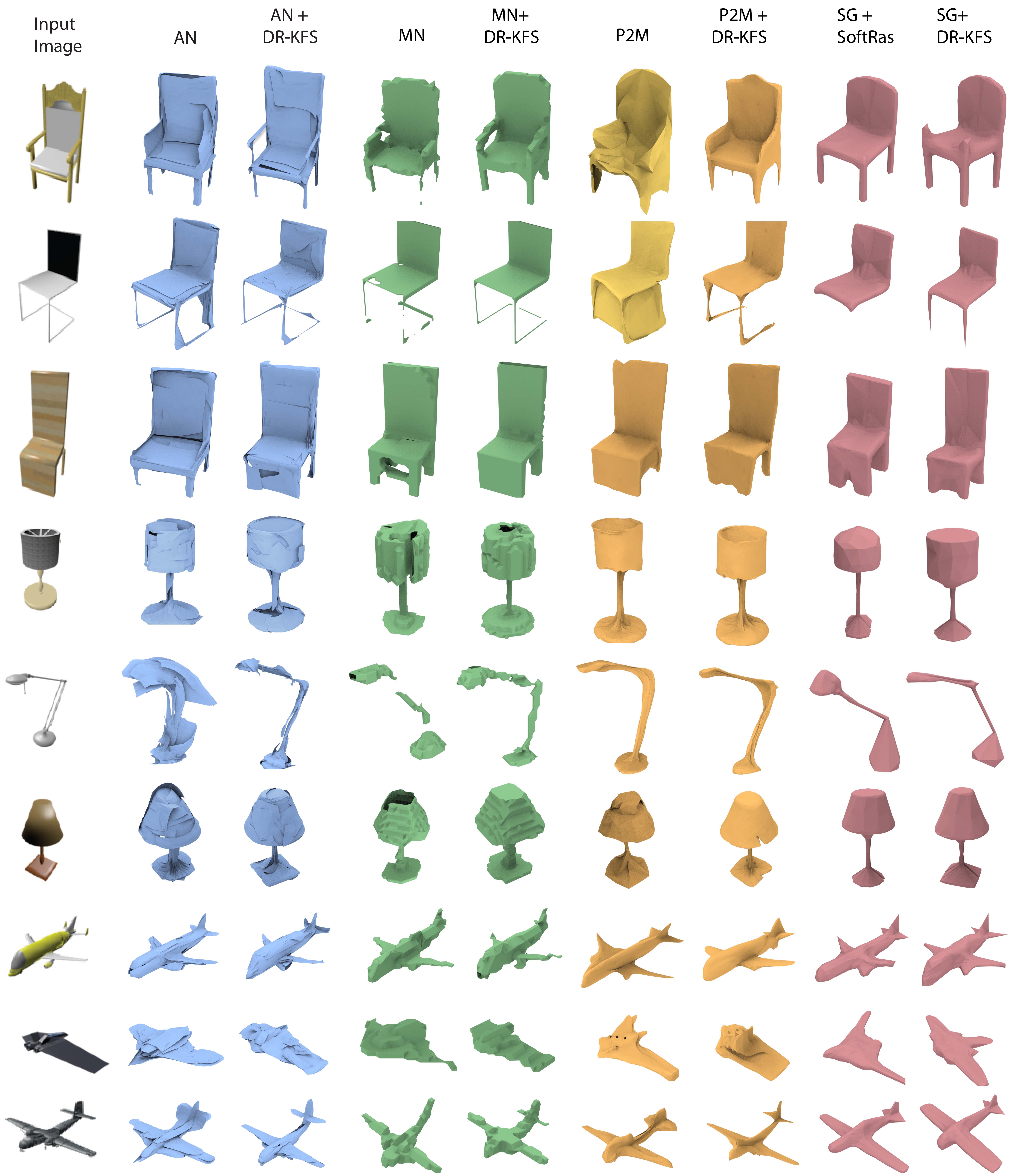}}
	\caption{A gallery of reconstructed 3D models obtained from AtlasNet (AN) \cite{atlasnet}, Matryoshka Net (MN) \cite{Matryoshka}, Pixel2Mesh (P2M) \cite{pixel2mesh} and Shape Generator (SG) \cite{softras}, trained using the metrics as adopted in the respective works and by using our DR-KFS metric. Given an input image, replacing the original training loss with our DR-KFS loss results in an improvement in the visual quality of the reconstructed 3D models as shown above, and also supported by the numbers in Table \ref{tab:score_table}.}
	\label{fig:gallery}
\end{figure*}

We assess the impact of our differentiable visual 
similarity
metric, DR-KFS, both qualitatively and quantitatively, on state-of-the-art single-view 3D reconstruction networks. A gallery of qualitative comparison on results generated when such networks are trained using DR-KFS is shown in Figure \ref{fig:gallery}. 

\subsection{Quantitative evaluation}
\label{quant_eval}

We consider existing single-view 3D reconstruction networks, including 3D-R2N2 \cite{3DR2N2}, OGN \cite{OGN}, AtlasNet \cite{atlasnet}, Matryoshka Network \cite{Matryoshka}, Pixel2Mesh \cite{pixel2mesh}, NMR \cite{NMR} and SoftRas\cite{softras} to quantify both (object-space)
shape distortion (CD and IoU) and the visual quality of the reconstructed models. The networks are trained either using 
their original losses or our metric DR-KFS. For visual quality measures, we choose representative tools including Shape
Google, Normal Consistency, F-score, LFD, and MVCNN.

Note that the shape generator used in SoftRas\cite{softras} and NMR\cite{NMR} are identical. We integrate DR-KFS training loss into this shape generator. Our method works well with mesh inputs. However, for voxel-based methods such as OGN/MN, we adopted a differentiable marching cubes layer (DMCL) proposed in \cite{DMCL} to convert voxels to meshes.


\vspace{-5pt}

\paragraph{Dataset.}
All the tested networks are trained and evaluated on the large-scale 3D CAD model dataset, ShapeNet \cite{ShapeNet}. We use four shape categories to train and test all the networks: airplane (4,045 models), car (3,533), chair (6,778), and lamp (2,318), with an 80-20 train-test split. Input data is prepared accordingly as consumed by each one of the aforementioned networks.

\vspace{-10pt}

\paragraph{Comparison results.}
Table~\ref{tab:score_table} shows the comparison results, where the numbers report {\em average\/} performance over all four object categories.
Note that IM-NET employs per-point reconstruction losses. 
In theory, DR-KFS can be employed for training IM-NET. However, the ensuing implementation requires an order of GPU memory more than what is currently available to us.
We report the IM-NET numbers for reference purposes only. 

Our first observation, perhaps a sanity check, is that when a network is trained using the new metric DR-KFS, it always performed better in DR-KFS, compared with the original network.
Second, when trained using DR-KFS, some of the networks even outperformed their counterparts trained with the original 
(object-space) losses, when the results are evaluated using CD and IoU --- this is the case 4 out of 10 or 40\% of the time.
In particular, Matryoshka was trained using IoU, but training with DR-KFS outperformed the original network when evaluated on IoU. Next, perhaps most importantly, we see that in majority of the cases (24 out of 30 or {\bf 80\%} of the time), replacing 
the original object-space losses using DR-KFS, the networks improved their performance in terms of visual shape similarity metrics.

\subsection{Qualitative evaluation}
\label{qual_eval}
To understand the visual quality of the reconstructed 3D models trained using DR-KFS as the network loss, we conduct two perceptual studies (PS) on Amazon Mechanical Turk (AMT), as described below.

\vspace{-5pt}

\paragraph{PS-1.}
In this study, we present 50 questions, each containing a single-view image and a pair of reconstructed models, one using the original loss and one using DR-KFS. Each model is rendered along two views, with one of the views aligned to the input image. We use the reconstructed models from AtlasNet \cite{atlasnet} and Matryoshka Net \cite{Matryoshka}. Models are selected at \emph{random} from the \emph{entire} reconstructed set. Turkers are presented with 50 questions in a random order, with randomized orderings of the reconstructed models as well as their view-images. 

\vspace{-5pt}

\paragraph{PS-2.}
We essentially repeat PS1, but instead of selecting the models from the entire reconstructed set, we select them \emph{randomly} from the \emph{top-20\%} of the reconstructed set, filtered based on the scores obtained using three visual similarity measures: LFD, ShapeGoogle and MvCNN.\\
Both PS1 and PS2 are forced-choice responses involving 80 different participants per study. For PS-1, DR-KFS metric received 71\% of the total votes (4000) and for PS-2, the percentage share shot up to 86.075\%, indicating a positive trend when high quality visual samples are considered for perceptual evaluation.
\vspace{-5pt}

\subsection{Design analysis}
\paragraph{Reconstruction using ``visual similarity loss vs.~image distortion loss"}

In Section \ref{sec:related}, we enlisted a sampler of representative works that address the task of single-view 3D reconstruction via image supervision. We argue that simply using a multi-view pipeline is insufficient in producing visually coherent results. To support this argument, we perform an experiment where we train the Shape Generator used in SoftRas \cite{softras} (state-of-the-art work for 3D reconstruction) using two different losses: (a) Image distortion loss (per-pixel MSE), and (b) Image-similarity loss (DR-KFS framework). We combine this analysis with the effect of visual-space representation described below.

\begin{figure}
\begin{floatrow}
\TopFloatBoxes
    \ffigbox
    {
    \includegraphics[width=0.9\linewidth]{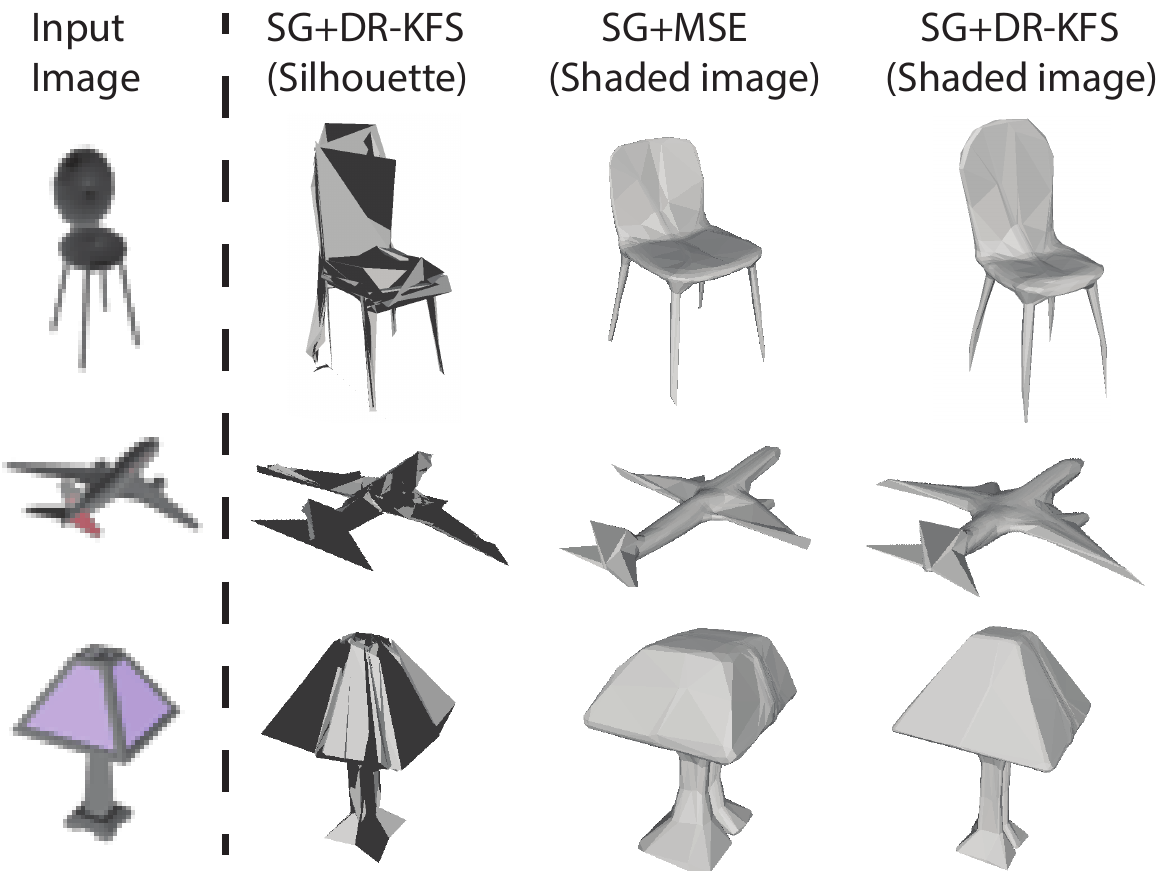}
    }
    {
    \caption{
    3D reconstruction results when the Shape Generator (SG) used in SoftRas \cite{softras} is trained using silhouettes vs.~shaded images corresponding to an input RGB image. 
    See Table \ref{tab:Ablation2} for the \emph{average} similarity scores for the shapes shown above. 
    }
    \label{fig:AB2}
    }
\capbtabbox
    {%
  \begin{tabular*}{\linewidth}{l|r|r}
		    \hline
         \multicolumn{1}{c|}{}& LFD & Shape Google \\
         \hline
         SG+DR-KFS &&\\
         (Silhouette)&6989&8.98\\
         \hline
         SG+MSE & &\\
         (Shaded Image) & 5156 & 7.33\\
         \hline
         SG+DR-KFS & &\\
         (Shaded Image) & \bf{4125}&\bf{5.48} \\
         \hline
		\end{tabular*}
	}
	{
    \caption{
    Training a simple Shape Generator (SG) adopted in SoftRas \cite{softras} with both DR-KFS and per-pixel MSE loss. DR-KFS, which performs image feature matching, essentially incorporates visual similarity, while per-pixel MSE is merely an image distortion loss on the image pixels. The reconstructed 3D models 
    are shown in Figure \ref{fig:AB2}.
    }
    \label{tab:Ablation2}
    }
\end{floatrow}
\vspace{-10 pt}
\end{figure}

\paragraph{Reconstruction from ``silhouette vs.~shaded images".}
Silhouette images only capture the boundary, without any surface information of the 3D shape. It is straightforward to infer that the reconstructed 3D shapes using silhouette images are likely to have bad surface quality (see Figure \ref{fig:AB2}), whereas using shaded images yields better reconstruction. To understand the differences in the reconstruction quality, we employ the shape generator used in SoftRas \cite{softras} for 3D reconstruction from either silhouette images or shaded images, with and without integrating DR-KFS pipeline.

The quantitative results for the effect of the two loss functions (visual similarity \emph{vs.} image distortion) are shown in Table \ref{tab:Ablation2}. DR-KFS, which measures image similarity (not distortion), wins over per-pixel MSE loss, when trained on shaded images, measured using visual shape similarity metrics such as LFD and Shape Google. Using per-pixel MSE loss on shaded images, however, is much better than using DR-KFS on silhouette images, indicating that shaded images are friendlier for the task of 3D reconstruction. This is also underscored visually in Figure \ref{fig:AB2}. This supports our claim that for a method to be \emph{truly visual}, it should not only make use of multi-view image pipeline, but should also employ a loss that captures visual similarity and not merely image distortion.

\begin{figure}
\begin{floatrow}
\TopFloatBoxes
    \ffigbox
    {
    \includegraphics[width=\linewidth]{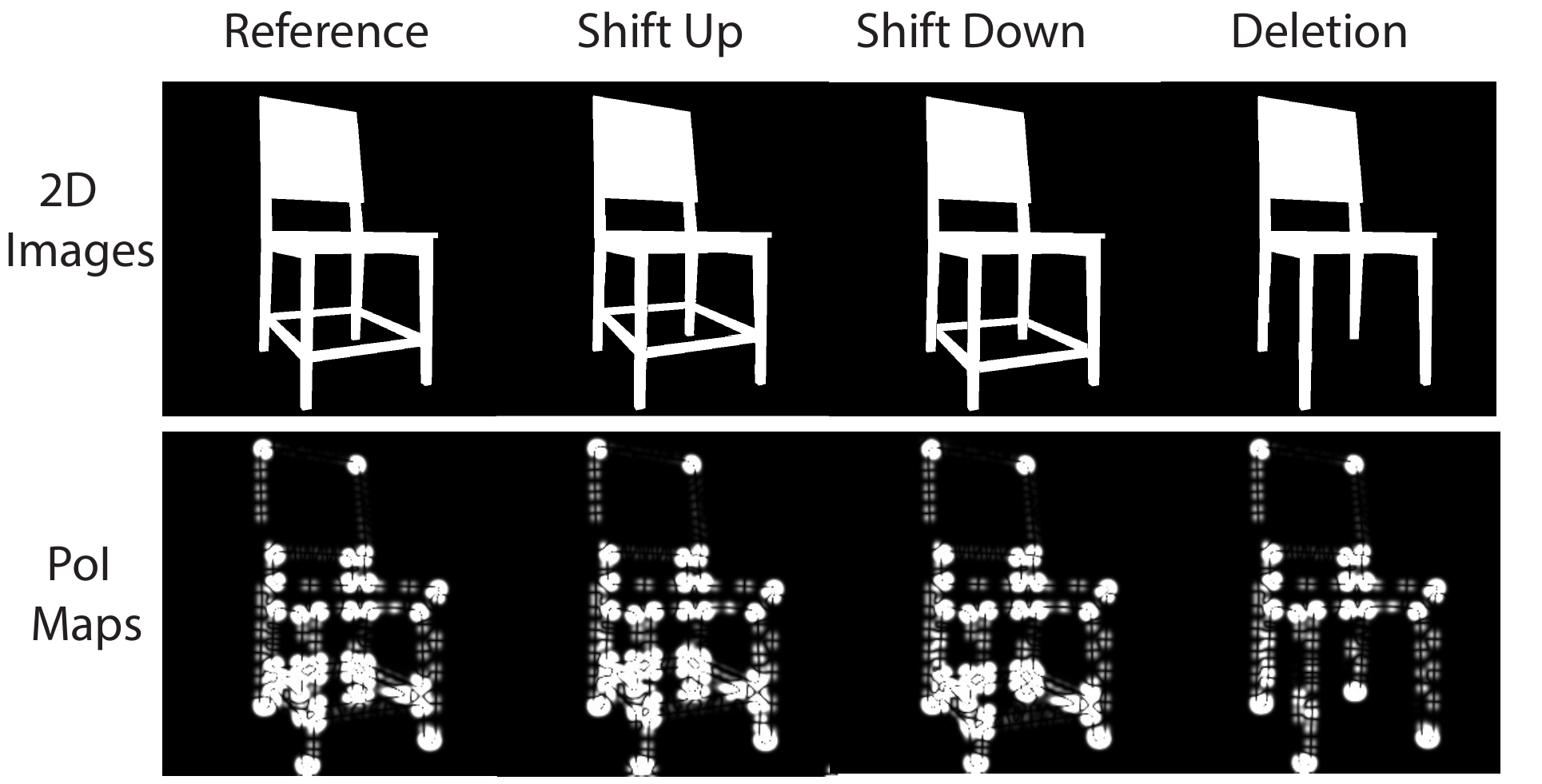}
    \vspace{-15 pt}
    }
    {
    \caption{Vertical positional shifts, and deletion of the leg bars. Binary images are shown on top and their corresponding PoI maps are on the bottom. Image similarity scores w.r.t the reference image for each operation using different image-level features are tabulated in Table \ref{tab:feat_space_scores}.}
    \label{fig:binary_images}
    }
\capbtabbox
    {%
  \begin{tabular*}{\linewidth}{l|r|r|r}
		    \hline
		    {Loss} &Shift Up & Shift down & Deletion \\
		    \hline
			MSE ($I$) &0.2961&\textit{0.3736}&0.2007\\
			\hline
			MSE&&&\\(LIFT feats) &0.1174&0.1231&\textit{0.1498}\\
			\hline
			MSE&&&\\(PoI Maps) &0.0137&\textit{0.0149}&0.0138\\
			\hline
			DR-KFS &0.0143&0.0160&\textit{0.0256}\\
			\hline
		\end{tabular*}
	}
	{
  \caption{Image similarity scores for the image-level operations shown in Fig \ref{fig:binary_images}, using: MSE loss on image pixels ($I$), image features using LIFT descriptors\cite{LIFT}, MSE loss on PoI map pixels, and DR-KFS framework. 
  Italicized numbers (row-wise) indicate sensitivity to the respective image-level operation.
  }%
  \label{tab:feat_space_scores}
    }
\end{floatrow}
\end{figure}

\paragraph{Features capturing perceptual differences.}
\label{essence}
Given a four-legged chair with a bar between adjacent pair of legs, humans can relate to it even after vertical shifts of the bars. However, if all the bars are removed, one cannot \emph{uniquely} associate the new chair to the original one (see Figure \ref{fig:binary_images}). Cognitively, the deletion operation is more visually apparent than the vertical positional shifts of the bars. It is desired that feature descriptors encompassing a visual similarity framework should be able to capture and reflect such perceptual differences. We investigate this by calculating the image-space distances on four different features of an image (rendered-view of a 3D shape): MSE on raw image pixels, MSE on  LIFT image features \cite{LIFT}, MSE on PoI maps, and DR-KFS approach. The input image and its corresponding PoI map are shown in Figure \ref{fig:binary_images}. In our experiment, vertical shifts of the bars are obtained by moving them five pixels up/down.
From Table \ref{tab:feat_space_scores}, we observe that DR-KFS (differentiable approach) and LIFT features (non-differentiable) are more sensitive towards the deletion operation compared to others. Moreover, image matching using DR-KFS and LIFT based local image features seem to be tolerant to small part shifts (2$^{nd}$ and 4$^{th}$ row in Table \ref{tab:feat_space_scores}), while per-pixel MSE on the image $I$ and the POI maps are quite sensitive to such relatively less perceptual changes than the deletion operation.

\section{Conclusion}
\label{sec:future}

We make a first step towards improving the visual quality of 3D reconstruction networks by 
making their training/reconstruction loss functions 
visual {\em and} driven by shape similarity rather than distortion
. Overall, the new differentiable 
visual similarity metric we develop, DR-KFS, is shown to improve the reconstruction quality for all
the tested networks (AtlasNet, Matryoshka, Pixel2Mesh, OGN, SoftRas, etc.), as judged by
a variety of visual similarity measures (LFD, MVCNN, Shape Google, etc.). This is clearly
a positive trend to motivate further investigation. However, the demonstrated advantage of DR-KFS is not yet ``across-the-board''. Our metric is still rather primitive as it does not utilize the most advanced and up-to-date tools that are available for multi-view rendering, feature extraction and learning, or image-space/perceptual assessment.

In general, the question of what the best 3D reconstruction is should depend on the application. 
For example, if the goal is to recognize or classify the shape, then visual similarity is
more important. As well, functional understanding of an acquired shape hinges on accurate 
recovery of shape structures, which would support the use of visual 
similarity
On the other hand, if the reconstructed 3D shape is to reflect accurate physical measures, then
object-space metrics should play a more prominent role.

One of the most obvious limitations of any visual 
similarity
DR-KFS included, is that it is oblivious to errors that are {\em hidden\/} from the viewers due to
occlusion. Such errors are not visible on the projected images, but they can be captured by
object-space shape distances. One possibility to explore is to combine differentiable object- and 
image-space shape metrics as the training loss, e.g., as a weighted sum which would preserve the 
differentiability. This could be a good strategy to address the distortion-perception 
trade-off~\cite{blau2018perception}, but leaves the question of how to choose the weight.

%
%
\bibliographystyle{splncs04}
\bibliography{egbib}
\end{document}